\documentclass[twocolumn]{article}
\usepackage{mathtools}
\usepackage{algorithm}
\usepackage{tabularx} 
\usepackage{array}
\usepackage{longtable}
\usepackage{balance}
\usepackage{xspace}
\newcolumntype{C}[1]{>{\centering\let\newline\arraybackslash\hspace{0pt}}m{#1}}
\usepackage{graphicx} 
\usepackage{multirow}
\usepackage{booktabs}
\usepackage{float}
\usepackage{amsmath,amssymb,amsthm,amsfonts,xcolor, color,stmaryrd}
\usepackage{enumerate}
\usepackage[utf8]{inputenc}
\usepackage[english]{babel}
\usepackage{tikz}
\usetikzlibrary{matrix, shapes, arrows, positioning, chains}
\usepackage{ragged2e}
\usepackage{etoolbox}
\usepackage{lipsum}
\usepackage[numbers,sort&compress]{natbib}
\usepackage{hyperref}
\usepackage{siunitx}
\usepackage{authblk}
\usepackage[margin=0.9in]{geometry}
\usepackage[english]{babel}
\date{}
\usepackage{textgreek} 
\apptocmd{\frame}{}{\justifying}{} 

\begin{document}

\title{\textbf{Cosmological Dynamics and Observational Constraints of a Logarithmic Herglotz $f(R,T)$ Gravity}}

\author[1]{Vishal M C \thanks{vishal.mc2023@vitstudent.ac.in}} 
\author[2]{Sankarsan Tarai \thanks{sankarsan.tarai@vit.ac.in}} 

\affil[1]{\it Department of Physics, School of Advanced Sciences, Vellore Institute of Technology, Chennai-600127, India} 

\affil[2]{\it Department of Mathematics, School of Advanced Sciences, Vellore Institute of Technology, Chennai-600127, India}

\date{}

\maketitle

\begin{abstract}
We investigate the cosmological dynamics and observational viability of a logarithmic Herglotz-type $f(R,T)$ gravity model described by
$f(R,T)=R+\beta\ln(-T/T_0)$.
Assuming a spatially flat Friedmann--Lema\^{i}tre--Robertson--Walker background and pressureless matter, the modified field equations are formulated as a coupled dynamical system involving the Herglotz field and the normalized matter density. A phenomenological closure relation is adopted for the effective Herglotz sector, allowing the background evolution to be determined numerically. The model is constrained using Cosmic Chronometer measurements, the DESI DR2 baryon acoustic oscillation compilation, and the Union3 supernova sample, considering both individual and combined datasets within a Bayesian Markov-chain Monte Carlo framework. The combined analysis provides a statistically consistent description of the background expansion. The reconstructed expansion history yields a present-day effective equation of state $w_{\rm eff,0}\simeq-0.702$ and exhibits the expected transition from accelerated expansion at late times toward decelerated expansion at higher redshifts. We further investigate the $Om(z)$ diagnostic and statefinder parameter. These results indicate that the logarithmic Herglotz model provides a consistent phenomenological description of the homogeneous and isotropic late-time expansion history, while allowing departures from the standard $\Lambda$CDM background.

\medskip


\end{abstract}

\section{Introduction}
\label{sec:introduction}

The discovery of the late-time accelerated expansion of the Universe through observations of Type Ia supernovae and other cosmological probes has fundamentally reshaped our understanding of cosmic evolution \cite{SupernovaSearchTeam:1997sck,SupernovaSearchTeam:1998cav,SupernovaCosmologyProject:1996grv,SupernovaCosmologyProject:1997zqe,SupernovaCosmologyProject:1998vns,SupernovaSearchTeam:1998fmf,Riess:2000yp,SupernovaSearchTeam:2004lze,SupernovaSearchTeam:1998bnz,SupernovaSearchTeam:2003cyd,WMAP:2003ivt,WMAP:2003elm,SDSS:2003eyi,HighZSNSearch:2005xhg,Boomerang:2000efg}. Within General Relativity (GR), a universe dominated by ordinary matter is expected to undergo decelerated expansion. However, observations \cite{SupernovaSearchTeam:1997sck,SupernovaSearchTeam:1998cav,SupernovaCosmologyProject:1996grv,SupernovaCosmologyProject:1997zqe,SupernovaCosmologyProject:1998vns,SupernovaSearchTeam:1998fmf,Riess:2000yp,SupernovaSearchTeam:2004lze,SupernovaSearchTeam:1998bnz,SupernovaSearchTeam:2003cyd,WMAP:2003ivt,WMAP:2003elm,SDSS:2003eyi,HighZSNSearch:2005xhg,Boomerang:2000efg} indicate a transition from decelerated to accelerated expansion at late times starting around redshift z = 1. In the standard $\Lambda$CDM framework, this acceleration is attributed to dark energy, which contributes approximately 70\% of the present cosmic energy density and is characterized by negative effective pressure. The nature of dark energy and the underlying mechanism driving cosmic acceleration remain open questions, motivating investigations beyond the standard cosmological paradigm.\\

In their conventional form, the Einstein field equations cannot adequately account for the observed late-time accelerated expansion of the Universe, as a pressureless matter-dominated cosmology corresponds to an equation-of-state parameter $w=0$ \cite{Copeland:2006wr}. Current cosmological studies primarily focus on constraining the equation-of-state parameter, $\omega$, which characterizes the dynamical properties of dark energy \cite{SNLS:2005qlf,WMAP:2006bqn,Riess:2006fw}. The observed late-time acceleration of the Universe requires an effective equation-of-state parameter satisfying $\omega < -1/3$. Such accelerated expansion can be realized by considering an exotic cosmic fluid with sufficiently negative pressure, implying that dark energy dominates the present cosmic energy budget \cite{Copeland:2006wr}. Alternatively, the cosmological constant, characterized by $\omega=-1$, provides the simplest explanation for the observed accelerated expansion within the standard $\Lambda$CDM framework \cite{Copeland:2006wr}. The $\Lambda$CDM model provides an excellent description of the observed late-time Universe \cite{Planck:2018nkj}. However, the cosmological constant $\Lambda$ is associated with the well-known fine-tuning problem, which motivates the exploration of alternative explanations for cosmic acceleration \cite{Weinberg:1988cp}.\\

Various cosmological and gravitational applications of $f(R)$ gravity, including inflation, dark energy, local gravity constraints, cosmological perturbations, and spherically symmetric solutions in weak and strong gravitational regimes, have been extensively reviewed in Ref.~\cite{DeFelice:2010aj}. Several comprehensive studies \cite{Nojiri:2017ncd,Capozziello:2011et,Clifton:2011jh,Sotiriou:2008rp} have reviewed recent developments in modified gravity and their cosmological applications, with particular emphasis on inflation, bouncing cosmologies, and late-time cosmic acceleration within $F(R)$, $F(G)$, and $F(T)$ gravity frameworks. Higher-order gravity theories have also been investigated in the context of quintessence and cosmic acceleration \cite{Capozziello:2002rd}, while dynamical dark-energy models have been extensively reviewed in \cite{Copeland:2006wr}. Furthermore, modified gravity models have been explored as a unified framework for describing both inflationary epoch \cite{Nojiri:2003ft} and the late-time cosmic acceleration \cite{Oikonomou:2020qah}, and viable $F(R)$ dark-energy models have been systematically examined in \cite{Oikonomou:2022wuk}.\\

Harko et al.~\cite{Harko:2011} proposed a generalization of $f(R)$ gravity by incorporating the trace $T$ of the energy-momentum tensor $T_{ij}$, leading to the formulation known as $f(R,T)$ gravity.
 It has been studied with reference to inflation \cite{Yeasmin:2022bqq,Gamonal:2020itt,Deb:2022hna,Chen:2022dyq}, dark energy \cite{Bhattacharjee:2019oim,Houndjo:2011fb,Pasqua:2013lha,Singh:2014kca,Reddy:2013vz}, dark matter \cite{Zaregonbadi:2016xna}, wormhole \cite{Godani:2020ndt,Shweta:2020fxs,Elizalde:2018arz,Moraes:2017rrv,Elizalde:2018frj,Moraes:2017dbs,Yousaf:2020srr}, pulsar \cite{dosSantos:2018nmu,Moraes:2015uxq}, white
dwarfs \cite{Rocha:2019nze}, gravitational waves \cite{Sharif:2019hyl,Bhatti:2020rzr}, scalar field models \cite{Singh:2014rwa,Goncalves:2021vci}, anisotropic models \cite{Rao:2015bga,Mishra:2017ycy,Mishra:2016hxb,Shamir:2014aea,Mishra:2017sdq}, bouncing cosmology \cite{Singh2018,Bhattacharjee:2020eec,Malik:2024dwd}, big-bang neucleosysnthesis \cite{Bhattacharjee:2020uhs} etc.\\

Despite its extensive applications, $f(R,T)$ gravity is subject to certain theoretical challenges. Owing to the non-minimal coupling between geometry and matter, the energy-momentum tensor is generally not covariantly conserved, such that $\nabla^{\mu}T_{\mu\nu}\neq 0$. Conservation can, however, be recovered for specific choices of the functional form of $f(R,T)$ \cite{PhysRevD.107.124005,dosSantos:2018nmu}. This non-conservation has traditionally been interpreted either as a theoretical drawback, associated with the violation of the equivalence principle through the emergence of an additional force on test particles, or as a possible manifestation of particle creation and annihilation processes \cite{Harko:2014pqa,Harko:2015pma,Pinto:2022tlu}.

A further, and perhaps more concrete, shortcoming concerns the cosmological viability of
the simplest linear extension of the theory, $f(R,T)=R+\alpha T$. It has been shown that,
within the standard Hamiltonian (Hilbert-Einstein type) variational formulation, this
model leads to a Hubble function of the form $H(z)\propto (1+z)^{B}$, with $B$ a constant
depending only on $\alpha$, which in turn implies a \emph{constant} deceleration parameter
$q=B-1$ \cite{Velten:2017hhf}. Such a model is therefore unable to reproduce the observed transition from a decelerated to an accelerated phase of cosmic expansion, and has consequently been dismissed as cosmologically inviable \cite{Velten:2017hhf}. This
difficulty, together with the generic non-conservation of $T_{\mu\nu}$, suggests that the standard variational treatment of $f(R,T)$ gravity may be missing an essential physical
ingredient.\\

A natural way to address these issues is to reinterpret the non-conservation of the
energy-momentum tensor not as a pathology, but as the signature of an effectively
dissipative process taking place within the gravitational sector itself. Dissipative
phenomena are ubiquitous in physical systems, ranging from bulk and shear viscosity in
astrophysical and cosmological fluids \cite{Maartens:1996vi} to irreversible particle
production in an expanding universe, and even to genuinely gravitational sources of irreversibility, such as gravitational wave emission and the coupling of the gravitational field to quantum matter \cite{Herrera:2024hby,Galley:2023byb}. However, the standard Hamilton variational principle, on which both General Relativity and $f(R,T)$ gravity are built, is known to be intrinsically unable to produce equations of motion containing first-order (dissipative) terms directly from a Lagrangian of the usual kinetic-minus-potential form \cite{Bauer:1931}. This motivates the search for an alternative variational framework capable of incorporating dissipation in a systematic and covariant manner.\\

Such a framework was provided, in a classical mechanical setting, by Herglotz already in
1930 \cite{Herglotz:1930}, who proposed a generalization of Hamilton's principle in which
the Lagrangian is allowed to depend explicitly on the action itself, rather than being a
function of the generalized coordinates and velocities alone. This seemingly modest
modification is sufficient to generate equations of motion containing genuine
dissipative (first-order derivative) terms, as is manifest, for instance, in the case of
the damped harmonic oscillator. A fully covariant formulation of the Herglotz principle,
suitable for application to field theories, was obtained only much more recently
\cite{Lazo:2018}, and its application to General Relativity was developed in
\cite{Lazo:2017,Paiva:2022}, leading to what is now known as Lazo's non-conservative
gravity. In this approach, the gravitational action is supplemented by an additional
one-form $\lambda_{\mu}$, which plays the role of a generalized, non-dynamical
dissipative coefficient, and which reduces, in a cosmological setting, to a single time-dependent function $\phi(t)$ \cite{Fabris:2017msx}. It was shown that the resulting background cosmology is formally equivalent to that of bulk viscous cosmology \cite{Fabris:2017msx}, and that the theory admits, without invoking an explicit dark energy component, a late-time accelerated phase of cosmic expansion \cite{Lazo:2017}.

Motivated by these considerations, and by the natural interpretation of the
non-conservation of $T_{\mu\nu}$ in $f(R,T)$ gravity as an effective manifestation of
dissipation, Wazny et al.~\cite{Wazny:2025jth} recently proposed a Herglotz-type
formulation of $f(R,T)$ gravity, obtained by applying the covariant Herglotz variational
principle to the $f(R,T)$ gravitational action. The resulting field equations extend
those of standard $f(R,T)$ gravity through the addition of a Herglotz tensor
$H_{\mu\nu}$, built out of the one-form $\lambda_{\mu}$, and reduce to the usual
$f(R,T)$ field equations, to Lazo's non-conservative gravity, and to the Einstein field
equations in the appropriate limits. In the weak-field, slow-motion regime, the Herglotz
contribution was shown to modify the Newtonian gravitational potential, allowing the
Herglotz vector to be observationally constrained through the perihelion precession of
Mercury and the relativistic deflection of light; remarkably, the resulting
frequency-dependent correction to the light-bending angle was found to reproduce the
scaling law inferred from Cassini spacecraft observations of light propagation through
the solar plasma \cite{Bertotti:2003rm}. At the cosmological level, two representative
models, $f(R,T)=R+\alpha T$ and $f(R,T)=R+\alpha T^{-1}$, were analyzed within this
Herglotz framework. In both cases, the Herglotz vector reduces to a single scalar
function of cosmic time, which, under suitable conditions, can play the role of an
effective cosmological constant, thereby offering an alternative route to late-time
cosmic acceleration. Most notably, it was shown that the linear model
$f(R,T)=R+\alpha T$, previously excluded in the standard Hamiltonian formulation on
account of its constant deceleration parameter \cite{Velten:2017hhf}, becomes observationally viable once formulated through the Herglotz variational approach, being able to reproduce a redshift-dependent deceleration parameter, as well as the $\Lambda$CDM behaviour of the Hubble function.\\

These results indicate that the Herglotz variational principle provides a promising and
mathematically consistent avenue for extending $f(R,T)$ gravity, curing at least one of
its known cosmological shortcomings while retaining a clear physical interpretation in
terms of dissipative, non-conservative dynamics. This raises the natural question of
whether other functional forms of $f(R,T)$, beyond the linear and simple inverse cases
already considered, can likewise benefit from this Herglotz extension, and whether such
models can be brought into quantitative agreement with a broader range of observational
data. In the present work, we address this question by considering a logarithmic
extension of $f(R,T)$ gravity, $f(R,T) =  R + \beta \ln\left(-\frac{T}{T_0}\right)$, formulated within the Herglotz variational framework. We derive the corresponding generalized Friedmann equations and constrain the model parameters using Cosmic Chronometers (CC), baryon acoustic oscillations (BAO), the Union3 supernova compilation, and their combined
dataset. We further assess the cosmological viability of the model through a set of cosmographic diagnostics, including the deceleration, statefinder as well as the $Om(z)$ diagnostic. Our results indicate that the logarithmic model, when formulated within the Herglotz
variational framework, provides a viable description of the late-time cosmic evolution and remains consistent with the $\Lambda$CDM cosmological paradigm..\\

The paper is organized as follows. Section~\ref{sec:herglotz} reviews
the Herglotz-type $f(R,T)$ framework and presents the corresponding gravitational field equations. Section~\ref{sec:cosmology} specializes the theory to a spatially flat FLRW spacetime and derives the background cosmological equations. The logarithmic functional form is introduced and analyzed in Section~\ref{subsec:logarithmic}, Section \ref{sec:observations} presents the observational datasets, likelihood construction, and Bayesian parameter estimation, including the joint CC+BAO+Union3 analysis. Section~\ref{sec:Diagnostics} investigates the reconstructed cosmological dynamics through $q(z)$, $Om(z)$, $w_{\rm eff}(z)$ and the statefinder diagnostic. Finally, Section~\ref{sec:conclusion} summarizes the principal results and discusses possible directions for future investigation.

\section{Herglotz-type $f(R,T)$ gravity}
\label{sec:herglotz}

The Herglotz variational framework provides a covariant formulation of
non-conservative gravitational dynamics by introducing an additional
one-form into the variational description. In the present work, we use
the formulation developed for Herglotz-type $f(R,T)$ gravity
\cite{Fabris:2017msx,Paiva:2022,Wazny:2025jth}. The additional one-form
$\lambda_\mu$ is assumed to be closed,
\begin{equation}
    \nabla_\mu \lambda_\nu-\nabla_\nu\lambda_\mu=0,
    \label{eq:lambda_closed}
\end{equation}
and can therefore be written locally as
\begin{equation}
    \lambda_\mu=\partial_\mu\chi ,
    \label{eq:lambda_gradient}
\end{equation}
where $\chi$ is a scalar function. The Herglotz one-form is treated as
a prescribed background quantity rather than as an independently
dynamical field \cite{Paiva:2022,Wazny:2025jth}.

For the Herglotz extension of $f(R,T)$ gravity, the relevant
Lagrangian density can be written schematically as
\begin{equation}
    \mathcal{L}
    =
    f(R,T)+F\mathcal{L}_{m}
    +\lambda_\mu s^\mu ,
    \label{eq:herglotz_lagrangian}
\end{equation}
where $R$ is the Ricci scalar, $\mathcal{L}_{m}$ denotes the matter
Lagrangian, $T$ is the trace of the energy--momentum tensor, and
$s^\mu$ is the Herglotz action-density vector. The coupling function
$F$ is taken to be constant and, in the cosmological application
considered below, we adopt $F=16\pi$ \cite{Wazny:2025jth}.

The energy--momentum tensor is defined by
\begin{equation}
    T_{\mu\nu}
    =
    -\frac{2}{\sqrt{-g}}
    \frac{\delta\left(\sqrt{-g}\mathcal{L}_{m}\right)}
    {\delta g^{\mu\nu}},
    \label{eq:Tmunu}
\end{equation}
with its trace given by
\begin{equation}
    T=g^{\mu\nu}T_{\mu\nu}.
    \label{eq:Ttrace}
\end{equation}
The metric dependence of $T_{\mu\nu}$ introduces the tensor
$\Theta_{\mu\nu}$, defined as
\begin{equation}
    \Theta_{\mu\nu}
    =
    g^{\alpha\beta}
    \frac{\delta T_{\alpha\beta}}
    {\delta g^{\mu\nu}}.
    \label{eq:Theta_definition}
\end{equation}
For a matter Lagrangian depending only on the metric, this tensor takes
the form
\begin{equation}
    \Theta_{\mu\nu}
    =
    -2T_{\mu\nu}
    +g_{\mu\nu}\mathcal{L}_{m}
    -2g^{\alpha\beta}
    \frac{\partial^{2}\mathcal{L}_{m}}
    {\partial g^{\mu\nu}\partial g^{\alpha\beta}} .
    \label{eq:Theta_general}
\end{equation}
These definitions follow the standard treatment of $f(R,T)$ gravity and its Herglotz extension \cite{Harko:2011,Wazny:2025jth}.

Defining
\begin{equation}
    f_R\equiv\frac{\partial f}{\partial R},
    \qquad
    f_T\equiv\frac{\partial f}{\partial T},
    \label{eq:fderivatives}
\end{equation}
the gravitational field equations obtained from the Herglotz
variational formulation are
\begin{equation}
\begin{aligned}
    f_R R_{\mu\nu}
    -\frac{1}{2}f g_{\mu\nu}
    &+\left(g_{\mu\nu}\Box-\nabla_\mu\nabla_\nu\right)f_R \\
    &+H_{\mu\nu}
    =
    \frac{F}{2}T_{\mu\nu}
    -f_T\left(T_{\mu\nu}+\Theta_{\mu\nu}\right).
\end{aligned}
\label{eq:herglotz_field_equations}
\end{equation}

where $\Box\equiv\nabla^\alpha\nabla_\alpha$ and the additional
Herglotz contribution is
\begin{equation}
    H_{\mu\nu}
    =
    f_R K_{\mu\nu}
    +\lambda_\mu\partial_\nu f_R
    +\lambda_\nu\partial_\mu f_R
    -2g_{\mu\nu}\lambda^\rho\partial_\rho f_R .
    \label{eq:Hmunu}
\end{equation}
Here, the tensor $K_{\mu\nu}$ is given by
\begin{equation}
\begin{split}
    K_{\mu\nu}
    ={}&
    \frac{1}{2}
    \left(
        \nabla_\mu\lambda_\nu
        +\nabla_\nu\lambda_\mu
    \right)
    -\lambda_\mu\lambda_\nu
\\
    &-
    g_{\mu\nu}
    \left(
        \nabla_\rho\lambda^\rho
        -\lambda_\rho\lambda^\rho
    \right).
\end{split}
\label{eq:Kmunu}
\end{equation}

For the cosmological matter sector, we consider a perfect fluid,
whose energy--momentum tensor is written as
\begin{equation}
    T_{\mu\nu}
    =
    (\rho+p)u_\mu u_\nu
    +p g_{\mu\nu},
    \qquad
    u_\mu u^\mu=-1 ,
    \label{eq:perfect_fluid}
\end{equation}
where $\rho$ and $p$ denote the energy density and pressure,
respectively. Following the prescription adopted in the Herglotz
$f(R,T)$ framework, we choose
\begin{equation}
    \mathcal{L}_{m}=p .
    \label{eq:Lm_pressure}
\end{equation}
Consequently, the trace of the energy--momentum tensor becomes
\begin{equation}
    T=-\rho+3p .
    \label{eq:Tperfectfluid}
\end{equation}

Equations~\eqref{eq:herglotz_field_equations}--\eqref{eq:Tperfectfluid}
provide the general set of relations required for the cosmological
analysis. In the next section, we specify the functional form of
$f(R,T)$ and specialize the above equations to a homogeneous and
isotropic FLRW background.

\section{Cosmological dynamics and specific models}
\label{sec:cosmology}

We now specialize the general Herglotz-type $f(R,T)$ framework to a
homogeneous and isotropic cosmological background. We consider a
spatially flat FLRW spacetime,
\begin{equation}
    ds^2=-dt^2+a^2(t)\delta_{ij}dx^i dx^j,
    \label{eq:FLRW}
\end{equation}
where $a(t)$ is the scale factor and the Hubble parameter is defined
as
\begin{equation}
    H=\frac{\dot a}{a}.
    \label{eq:Hubble}
\end{equation}

Following the cosmological construction of Herglotz-type
$f(R,T)$ gravity \cite{Wazny:2025jth}, spatial homogeneity and isotropy
restrict the Herglotz one-form to the form
\begin{equation}
    \lambda_\mu=(\varphi(t),0,0,0),
    \label{eq:lambda_FLRW}
\end{equation}
where $\varphi(t)$ is a function of cosmic time. We assume that the
matter sector is described by a pressureless fluid,
\begin{equation}
    p=0,
    \qquad
    T=-\rho .
    \label{eq:dust}
\end{equation}

For the class of models considered in this work, we write
\begin{equation}
    f(R,T)=R+g(T).
    \label{eq:general_model}
\end{equation}
It follows immediately that
\begin{equation}
    f_R=1,
    \qquad
    \dot f_R=0,
    \qquad
    \ddot f_R=0.
    \label{eq:fR_constant}
\end{equation}
Consequently, the Herglotz contribution appearing in the general
field equations reduces to
\begin{equation}
    H_{\mu\nu}=K_{\mu\nu}.
    \label{eq:H_equals_K}
\end{equation}

For convenience, we define
\begin{equation}
    G(\rho)\equiv g(T)\big|_{T=-\rho}.
    \label{eq:G_definition}
\end{equation}
The cosmological field equations for the above class can then be
written as
\begin{equation}
    3H^2
    =
    3\varphi H
    +8\pi\rho
    -G'(\rho)\rho
    -\frac{1}{2}G(\rho),
    \label{eq:general_Friedmann}
\end{equation}
and
\begin{equation}
    \dot H
    =
    \frac{1}{2}\dot\varphi
    -\frac{1}{2}\varphi H
    -\frac{1}{2}\varphi^2
    -4\pi\rho
    +\frac{1}{2}G'(\rho)\rho .
    \label{eq:general_Hdot}
\end{equation}
These equations follow from the generalized FLRW equations of
Herglotz-type $f(R,T)$ gravity after imposing $f_R=1$ and the dust
condition. They provide the starting point for the specific
cosmological models considered below \cite{Wazny:2025jth}.

Since the Herglotz function $\varphi(t)$ does not possess an
independent equation of motion in this framework, an additional
closure relation is required. We adopt an effective equation of state
for the geometric/Herglotz sector,
\begin{equation}
    p_{\rm eff}=w\rho_{\rm eff},
    \label{eq:effective_EOS}
\end{equation}
where $w$ is treated as a constant parameter. The resulting evolution
equation for the Herglotz function is
\begin{equation}
    \dot\varphi
    =
    \varphi^2
    -(2+3w)\varphi H
    +\frac{1}{2}(1+w)G(\rho)
    +wG'(\rho)\rho .
    \label{eq:phi_general}
\end{equation}
To analyze the system numerically, it is convenient to introduce a set
of dimensionless variables,
\begin{equation}
    \tau=H_0 t,
    \qquad
    H=H_0 h,
    \qquad
    \varphi=H_0\Phi,
    \label{eq:dimensionless_variables}
\end{equation}
where $H_0$ denotes the present-day value of the Hubble parameter, and
$h(\tau)$ and $\Phi(\tau)$ are the dimensionless Hubble and Herglotz
functions, respectively. The matter energy density is likewise
normalized as
\begin{equation}
    \rho=\frac{3H_0^2}{8\pi}\,r ,
    \label{eq:normalized_density}
\end{equation}
where $r(\tau)$ is the dimensionless (normalized) matter density.
Equations~\eqref{eq:general_Friedmann},
\eqref{eq:general_Hdot}, and \eqref{eq:phi_general} constitute the
general dynamical system used for the Logarithmic model studied in this work.

\subsection{Logarithmic model}
\label{subsec:logarithmic}

As a second possibility, we consider a logarithmic dependence on the
trace of the energy-momentum tensor. Logarithmic forms of the
trace-dependent function $h(T)$ have previously been investigated in
$f(R,T)$ gravity, including in cosmological and observational
analyses \cite{Elizalde:2019ote,Sardar:2023iha}. Motivated by these studies,
we consider the logarithmic model
\begin{equation}
    f(R,T)
    =
    R+\beta\ln\left(-\frac{T}{T_0}\right),
    \label{eq:log_model}
\end{equation}
where $\beta$ is a constant parameter and $T_0$ is a reference
quantity with the same dimensions as $T$, introduced so that the
argument of the logarithm is dimensionless. For the pressureless matter source considered here, $T=-\rho$ also $-T>0$ and therefore
\begin{equation}
    G(\rho)
    =
    \beta\ln\left(\frac{\rho}{T_0}\right).
    \label{eq:G_log}
\end{equation}
The derivative is
\begin{equation}
    G'(\rho)=\frac{\beta}{\rho},
    \qquad
    G'(\rho)\rho=\beta .
    \label{eq:Gprime_log}
\end{equation}

Using the same dimensionless variables introduced in
Eqs.~\eqref{eq:dimensionless_variables}--\eqref{eq:normalized_density}
and choosing $T_0=\rho_{c0}$, we define
\begin{equation}
    B=\frac{\beta}{3H_0^2}.
    \label{eq:B_definition}
\end{equation}
The function $G(\rho)$ and its derivative contribution then become
\begin{equation}
    G(\rho)=3H_0^2B\ln r,
    \qquad
    G'(\rho)\rho=3H_0^2B .
    \label{eq:G_log_dimensionless}
\end{equation}

The dimensionless Friedmann constraint for the logarithmic model is
\begin{equation}
    h^2
    =
    \Phi h+r
    -
    B\left(1+\frac{1}{2}\ln r\right).
    \label{eq:log_constraint}
\end{equation}

The corresponding evolution equation for $h$ is
\begin{equation}
    h'
    =
    -\frac{3}{2}(1+w)\Phi h
    -\frac{3}{2}r
    +\frac{3}{2}(1+w)B
    \left(1+\frac{1}{2}\ln r\right).
    \label{eq:log_hprime}
\end{equation}

The Herglotz function evolves according to
\begin{equation}
    \Phi'
    =
    \Phi^2
    -(2+3w)\Phi h
    +\frac{3}{2}(1+w)B\ln r
    +3wB .
    \label{eq:log_Phiprime}
\end{equation}

Since the constraint
Eq.~\eqref{eq:log_constraint} does not provide an independent
algebraic evolution equation for $r$, we again differentiate it
implicitly. This gives
\begin{equation}
    r'
    =
    \frac{
        2hh'-\Phi'h-\Phi h'
    }{
        1-\dfrac{B}{2r}
    } .
    \label{eq:log_rprime}
\end{equation}

Hence, the logarithmic model is also described by a closed
first-order system for $\{h,\Phi,r\}$ through
Eqs.~\eqref{eq:log_hprime},
\eqref{eq:log_Phiprime}, and
\eqref{eq:log_rprime}.

\subsection{Redshift-space formulation}
\label{subsec:redshift}

For comparison with observational data, it is convenient to express
the dynamical system in terms of the cosmological redshift,
\begin{equation}
    1+z=\frac{1}{a},
\end{equation}
where the present scale factor is normalized to $a_0=1$. Since
\begin{equation}
    \frac{d}{d\tau}
    =
    -(1+z)h(z)\frac{d}{dz},
    \label{eq:tau_z_relation}
\end{equation}
the dimensionless Hubble function and the Herglotz function can be
evolved directly as functions of redshift.

For the logarithmic model, the corresponding equations are
\begin{equation}
    \frac{dh}{dz}
    =
    \frac{
        \frac{3}{2}(1+w)\Phi h
        +\frac{3}{2}r
        -\frac{3}{2}(1+w)B
        \left(1+\frac{1}{2}\ln r\right)
    }
    {(1+z)h},
\label{eq:log_hz}
\end{equation}

\begin{equation}
    \frac{d\Phi}{dz}
    =
    -\frac{
        \Phi^2
        -(2+3w)\Phi h
        +\frac{3}{2}(1+w)B\ln r
        +3wB
    }
    {(1+z)h},
\label{eq:log_Phi_z}
\end{equation}
and
\begin{equation}
    \frac{dr}{dz}
    =
    -\frac{1}{(1+z)h}
    \frac{
        2hh'-\Phi'h-\Phi h'
    }{
        1-\dfrac{B}{2r}
    } .
\label{eq:log_r_z}
\end{equation}

The present-day normalization of the Hubble function is
\begin{equation}
    h(0)=1,
    \label{eq:h_initial}
\end{equation}
while the present value of the Herglotz function is denoted by
\begin{equation}
    \Phi(0)=\Phi_0.
    \label{eq:Phi_initial}
\end{equation}
The present normalized matter density $r(0)$ is not treated as an
independent parameter. Instead, it is determined from the Friedmann
constraint, Eq.~\eqref{eq:log_constraint}, evaluated at $z=0$.

The observable Hubble parameter is finally obtained from
\begin{equation}
    H(z)=H_0 h(z).
    \label{eq:H_of_z}
\end{equation}
Therefore, once the model parameters and the initial conditions are
specified, the above system determines the theoretical expansion
history $H(z)$ that can be directly compared with cosmological
observations.

\section{Observational Constraints}
\label{sec:observations}
\subsection{Cosmic Chronometer Data}
\label{subsec:cc_constraints}

We first constrain the logarithmic Herglotz-type gravity model using the cosmic chronometer (CC) measurements of the Hubble parameter. We employ a compilation of 31 \(H(z)\) measurements spanning $0.07 \lesssim z \lesssim 2.34$ \cite{Guo:2015gpa}. Since the chronometer technique provides a direct estimate of the expansion rate through the differential aging of passively evolving galaxies, the CC dataset offers a direct probe of the background cosmological expansion without requiring a calibrated luminosity distance relation.

For a parameter vector $\boldsymbol{\theta}=(H_0,B,w,\Phi_0)$, the theoretical Hubble parameter $H_{\rm th}(z;\boldsymbol{\theta})$ is obtained by solving
the background equations of the logarithmic Herglotz model. In the numerical implementation, the dimensionless Hubble parameter $h(z)=H(z)/H_0$ and the Herglotz variable $\Phi(z)$ are evolved simultaneously, whereas the dimensionless matter density is determined
algebraically from the modified Friedmann constraint. This procedure ensures that the numerical evolution remains consistent with the constraint equation throughout the integration.\\

The parameter estimation is performed within a Bayesian framework
using the standard Gaussian likelihood constructed from the CC
measurements,
\begin{equation}
\chi^2_{\rm CC}(\boldsymbol{\theta})
=
\sum_{i=1}^{N_{\rm CC}}
\left[
\frac{H_{\rm CC}(z_i)
-
H_{\rm th}(z_i;\boldsymbol{\theta})}
{\sigma_{H,i}}
\right]^2,
\label{eq:chi2_cc}
\end{equation}
where $H_{\rm CC}(z_i)$ and $\sigma_{H,i}$ denote the observed Hubble
parameter and its corresponding uncertainty, respectively, and
$N_{\rm CC}=31$ is the number of CC data points. Flat priors are
adopted over the parameter ranges considered in the analysis. The
posterior distributions are subsequently sampled using the affine-
invariant Markov Chain Monte Carlo method implemented in the
\texttt{emcee} package \cite{Foreman-Mackey:2012any}.\\

\begin{figure}
    \centering
    \includegraphics[width=0.95\linewidth]{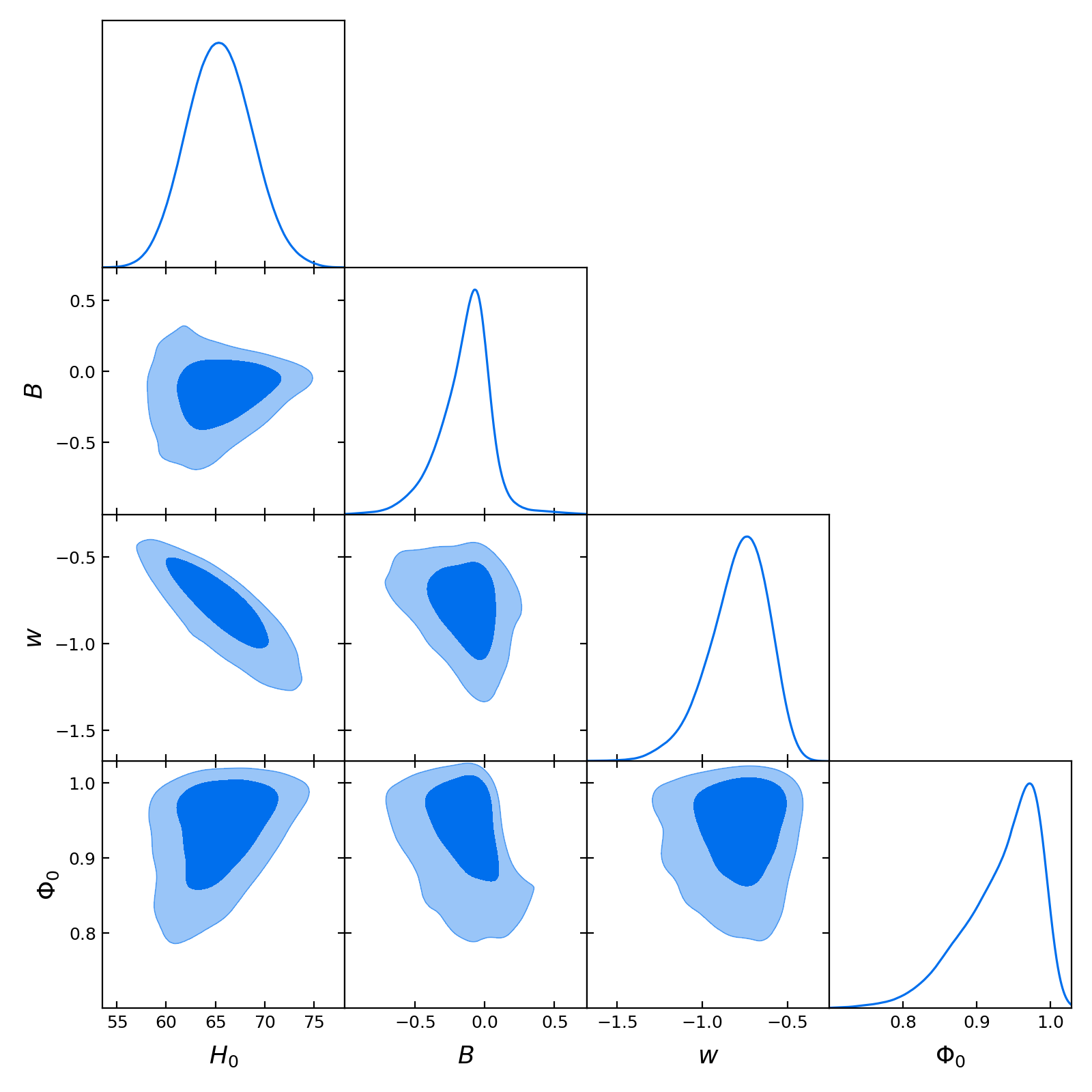}
    \caption{Marginalized one-dimensional and joint two-dimensional posterior distributions of the parameters $(H_0,B,w,\Phi_0)$ obtained from the cosmic chronometer data. The contours enclose the $68\%$ and $95\%$ credible regions.}
    \label{fig:CC_corner}
\end{figure}

The marginalized and joint posterior distributions obtained from the CC analysis are shown in Fig.~\ref{fig:CC_corner}. The diagonal panels
show the one-dimensional marginalized posterior distributions, while the off-diagonal panels display the corresponding two-dimensional joint posterior distributions. The four parameters are constrained to
finite posterior intervals within the adopted priors. The posterior of $H_0$ is localized around the inferred present-day expansion rate, while $B$ is constrained to negative values, indicating a preference for a negative coefficient of the logarithmic contribution. The
posterior of $w$ is concentrated around negative values, and $\Phi_0$ remains localized within the allowed parameter range.\\

\subsection{Baryon Acoustic Oscillation Data}
\label{subsec:bao_constraints}

Baryon Acoustic Oscillations (BAO) provide a powerful geometric probe of the expansion history of the Universe by measuring the characteristic scale imprinted in the spatial distribution of
matter. The BAO scale acts as a standard ruler and allows cosmological distances to be constrained at different effective redshifts. In this work, we use the BAO measurements from the second data release (DR2) of the Dark Energy Spectroscopic Instrument
(DESI) \cite{DESI:2025zgx,DESI:2025zpo}. The DESI DR2 BAO compilation combines measurements from different galaxy and quasar tracers together with
the high-redshift Ly$\alpha$ forest measurements, providing constraints on the expansion history over a broad redshift range.\\

For a given parameter set
\begin{equation}
\boldsymbol{\theta}
=
(H_0,B,w,\Phi_0,r_d),
\label{eq:bao_parameters}
\end{equation}
the background expansion is first obtained by solving the modified
cosmological equations of the logarithmic Herglotz model. The
dimensionless Hubble parameter $h(z)=H(z)/H_0$ and the Herglotz
variable $\Phi(z)$ are evolved numerically, while the dimensionless
matter density is determined from the algebraic Friedmann constraint.
The resulting continuous solution for $H(z)$ is then used to evaluate
the relevant BAO distance measures.

The transverse comoving distance is calculated according to
\begin{equation}
D_M(z)
=
c\int_0^z\frac{dz'}{H(z')},
\label{eq:DM}
\end{equation}
while the radial distance scale is given by
\begin{equation}
D_H(z)
=
\frac{c}{H(z)}.
\label{eq:DH}
\end{equation}
The volume-averaged distance entering the isotropic BAO measurements
is defined as
\begin{equation}
D_V(z)
=
\left[
zD_M^2(z)D_H(z)
\right]^{1/3}.
\label{eq:DV}
\end{equation}
The theoretical quantities compared with the DESI measurements are
therefore $D_M/r_d$, $D_H/r_d$, and $D_V/r_d$, where $r_d$ denotes the
sound horizon at the drag epoch. In the present analysis, $r_d$ is
treated as a free parameter and is constrained simultaneously with
the cosmological parameters.

The BAO likelihood is constructed using the full covariance matrix,
which is particularly important because the different BAO
observables are not statistically independent. Defining the residual
vector by
\begin{equation}
\Delta_i
=
X_i^{\rm obs}
-
X_i^{\rm th}(\boldsymbol{\theta}),
\label{eq:bao_residual}
\end{equation}
the corresponding chi-square statistic is
\begin{equation}
\chi^2_{\rm BAO}
=
\Delta_i
\left(C^{-1}\right)_{ij}
\Delta_j,
\label{eq:chi2_bao}
\end{equation}
where $C$ denotes the full covariance matrix of the BAO data vector.
The posterior probability is subsequently sampled using an
affine-invariant Markov Chain Monte Carlo analysis. Flat priors are
adopted for all five parameters within the ranges specified in the
analysis.\\

\begin{figure}
    \centering
    \includegraphics[width=0.95\linewidth]{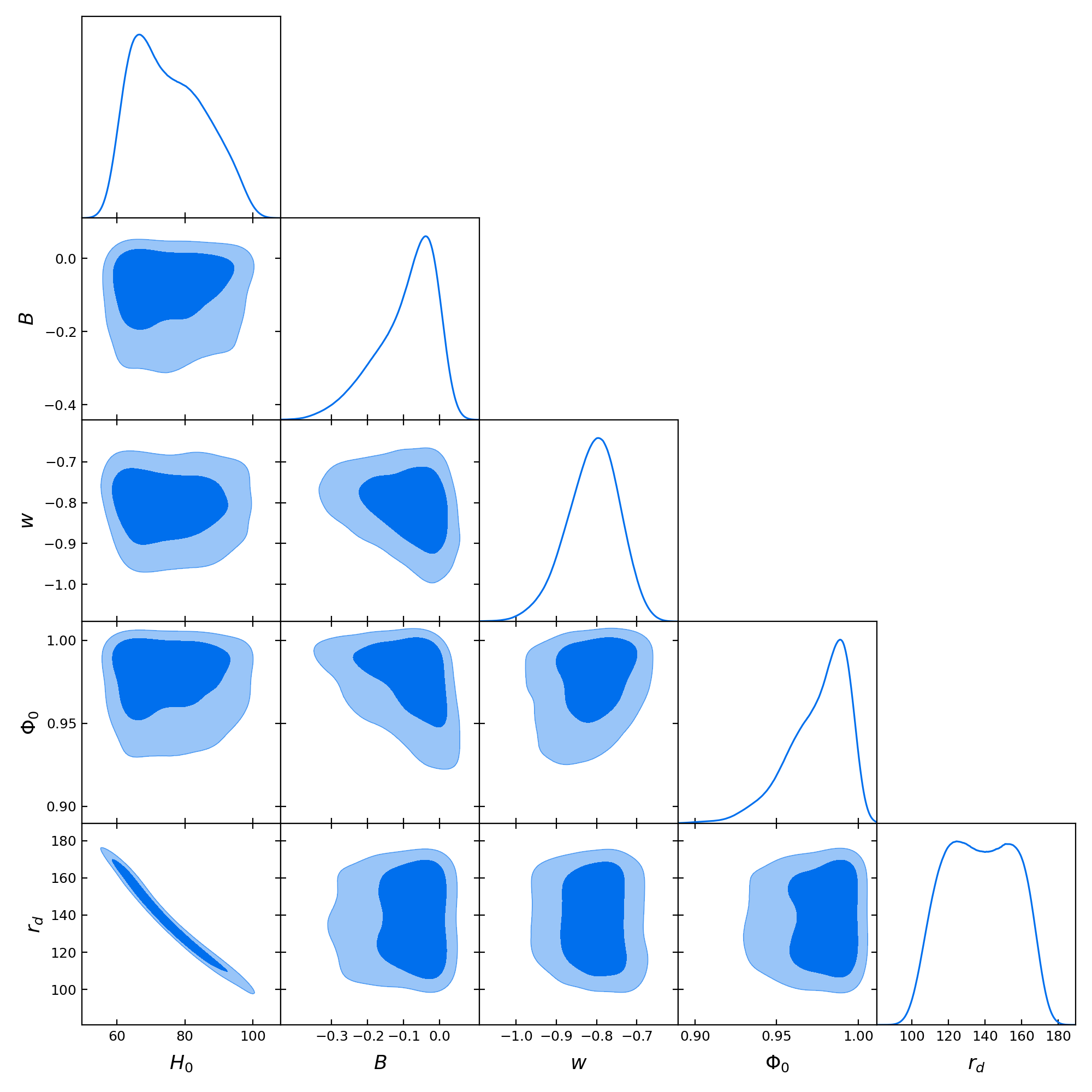}
    \caption{Marginalized one-dimensional and joint two-dimensional posterior distributions of the parameters $(H_0,B,w,\Phi_0,r_d)$ obtained from the DESI DR2 BAO data. The contours enclose the $68\%$ and $95\%$ credible regions.}
    \label{fig:BAO_corner}
\end{figure}

The resulting marginalized and joint posterior distributions are shown
in Fig.~\ref{fig:BAO_corner}. The diagonal panels show the
one-dimensional marginalized posterior distributions, while the
off-diagonal panels display the corresponding two-dimensional joint
posterior distributions. The BAO data constrain all five parameters
within the adopted prior ranges. The posterior of $H_0$ is localized
over a finite interval, while $B$ is constrained to negative values.
The posterior of $w$ is concentrated around negative values, and
$\Phi_0$ remains localized within the allowed parameter range. The
sound-horizon parameter $r_d$ is also constrained by the BAO
measurements and exhibits a strong correlation with $H_0$.

\subsection{Union3 Supernova Data}
\label{sec:union3}

We further test the logarithmic Herglotz-type gravity model against the Union3 compilation of Type Ia supernovae \citep{Rubin:2023jdq}. We use the compressed ``UNITY1.5'' release consisting of 22 redshift bins, together with the full inverse covariance matrix accounting for the statistical and systematic uncertainties. Unlike the cosmic chronometer and BAO observables, which constrain the expansion history through $H(z)$ and distance ratios, respectively, Type Ia supernovae probe the luminosity distance--redshift relation. The Union3 data therefore provide an independent constraint on the integrated expansion history.\\

The theoretical distance modulus is given by
\begin{equation}
    \mu_{\rm th}(z)
    =
    5\log_{10}
    \left[
    \frac{D_L(z)}{\mathrm{Mpc}}
    \right]+25,
    \label{eq:mu_theory}
\end{equation}
where, for a spatially flat background, the luminosity distance is
\begin{equation}
    D_L(z)
    =
    (1+z)c
    \int_0^z
    \frac{dz'}{H(z')}.
    \label{eq:luminosity_distance}
\end{equation}

For each point in the parameter space, the background expansion history $H(z)$ is obtained by numerically solving the coupled Friedmann--Herglotz equations. The dimensionless Hubble parameter $h(z)=H(z)/H_0$ and the Herglotz variable $\Phi(z)$ are evolved using the redshift-transformed background equations, while the dimensionless matter density $r(z)$ is determined algebraically from the modified Friedmann constraint. The resulting expansion history is then used to
evaluate the luminosity distance and the corresponding theoretical distance modulus.\\

The parameter vector considered in the Union3 analysis is
\begin{equation}
    \boldsymbol{\theta}
    =
    (H_0,B,w,\Phi_0),
    \label{eq:union3_parameters}
\end{equation}
where $H_0$ denotes the present-day Hubble parameter, $B$ controls the
strength of the logarithmic modification, $w$ characterizes the
effective Herglotz-sector closure, and $\Phi_0$ represents the present-day value of the Herglotz variable. Flat priors are adopted for all four
parameters within the ranges specified in the analysis.

The likelihood is constructed using the full covariance matrix of the Union3 distance-modulus data. Defining the residual vector as
\begin{equation}
    \Delta\mu_i
    =
    \mu_{\rm th}(z_i;\boldsymbol{\theta})
    -
    \mu_{\rm obs}(z_i),
    \label{eq:union3_residual}
\end{equation}
the corresponding chi-square statistic is
\begin{equation}
    \chi^2_{\rm Union3}
    =
    \Delta\mu_i
    \left(C^{-1}\right)_{ij}
    \Delta\mu_j,
    \label{eq:chi2_union3}
\end{equation}
where $C$ denotes the covariance matrix of the Union3 distance-modulus measurements. The use of the full covariance matrix accounts for the correlated statistical and systematic uncertainties in the supernova compilation.\\

\begin{figure}
    \centering
    \includegraphics[width=0.95\linewidth]{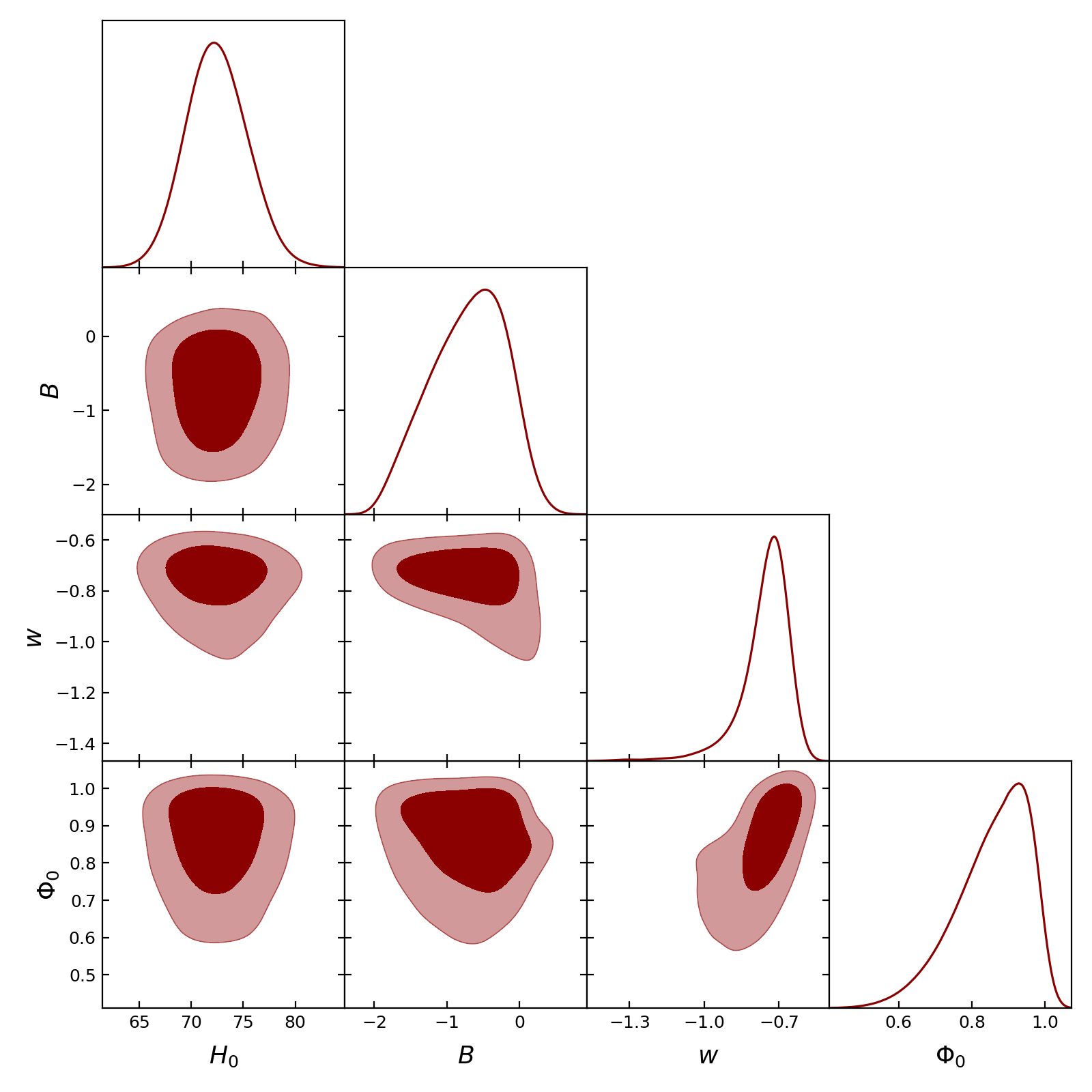}
    \caption{Marginalized one-dimensional and joint two-dimensional posterior distributions of the parameters $(H_0,B,w,\Phi_0)$ obtained from the Union3 data. The contours enclose the $68\%$ and $95\%$ credible regions.}
    \label{fig:Union3_corner}
\end{figure}

The resulting marginalized and joint posterior distributions are shown in Fig.~\ref{fig:Union3_corner}. The diagonal panels show the one-dimensional marginalized posterior distributions, while the off-diagonal panels display the corresponding two-dimensional joint posterior distributions. The posterior of $H_0$ is localized around $H_0\approx72~\mathrm{km\,s^{-1}\,Mpc^{-1}}$, while $B$ is preferentially constrained to negative values. The posterior of $w$ is concentrated around negative values, peaking near $w\approx-0.75$--$-0.8$. The parameter $\Phi_0$ is concentrated toward the upper part of the adopted prior range, with the posterior
approaching the boundary at $\Phi_0=1$.

\subsection{Combined Analysis}
\label{subsec:joint_constraints}

To obtain joint constraints on the logarithmic Herglotz-gravity model, we combine three complementary probes of the late-time expansion history: cosmic chronometer (CC) measurements of $H(z)$, the Union3 Type Ia supernova compilation, and DESI DR2 baryon acoustic oscillation (BAO) measurements.\\

For each parameter combination, the background evolution is obtained by numerically solving the coupled Friedmann--Herglotz equations derived in Sec.~\ref{subsec:redshift}. The dimensionless Hubble parameter $h(z)=H(z)/H_0$ and the Herglotz variable $\Phi(z)$ are evolved using the redshift-transformed background equations, while the dimensionless matter density $r(z)$ is determined algebraically from the modified Friedmann constraint. The resulting $H(z)$ is used consistently to evaluate the CC, supernova, and BAO observables. $r_d$ is treated as a free parameter. The joint parameter vector is therefore
\begin{equation}
    \boldsymbol{\theta}
    =
    (H_0,B,w,\Phi_0,r_d).
\end{equation}

The total chi-square is,
\begin{equation}
    \chi^2_{\rm tot}
    =
    \chi^2_{\rm CC}
    +
    \chi^2_{\rm Union3}
    +
    \chi^2_{\rm BAO}.
\end{equation}

Flat priors are adopted for the five parameters,
\begin{equation}
\begin{aligned}
    H_0 &\in [50,100]~\mathrm{km\,s^{-1}\,Mpc^{-1}},\\
    B &\in [-2,2],\\
    w &\in [-2,0],\\
    \Phi_0 &\in [-1,1],\\
    r_d &\in [100,200]~\mathrm{Mpc}.
\end{aligned}
\end{equation}

\begin{figure}
    \centering
    \includegraphics[width=0.95\linewidth]{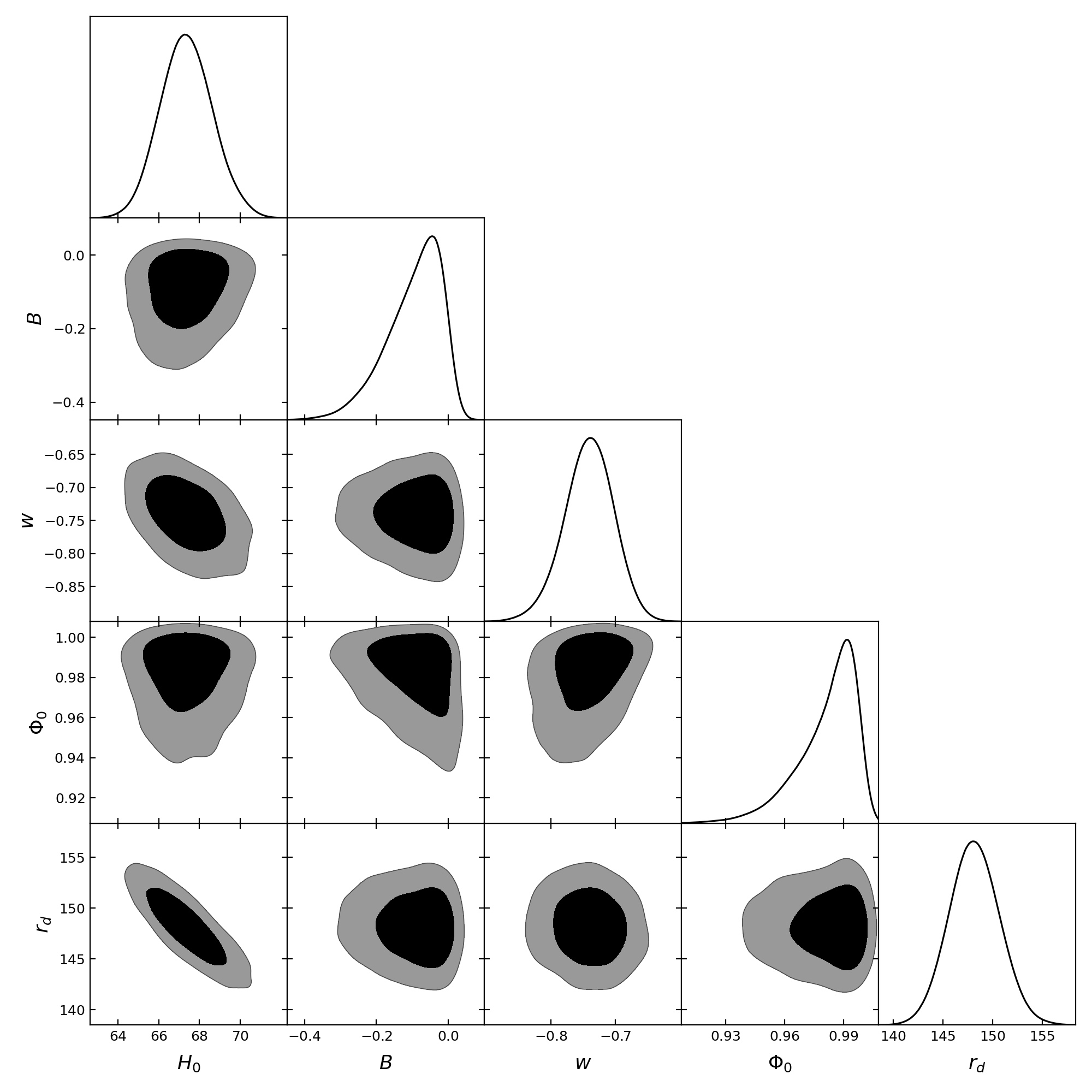}
    \caption{Marginalized one-dimensional and joint two-dimensional
    posterior distributions of the shared parameters
    $(H_0,B,w,\Phi_0,r_d)$ from the combined CC, Union3, and DESI DR2 BAO
    analysis. The contours enclose the $68\%$ and $95\%$ credible
    regions.}
    \label{fig:combined_corner}
\end{figure}

The joint posterior distributions are shown in Fig.~\ref{fig:combined_corner}. Combining all three probes tightens the constraints substantially relative
to any single dataset. The combined analysis yields
$H_0=67.3488^{+1.2937}_{-1.2868}\,
\mathrm{km\,s^{-1}\,Mpc^{-1}}$, with substantially tighter constraints than those obtained from the individual datasets. The logarithmic coupling parameter is constrained to $B=-0.0796^{+0.0628}_{-0.0933}$, indicating that the combined posterior is concentrated at negative values of $B$. The posterior remains sufficiently close to the standard-gravity limit to warrant further investigation of the statistical preference for the logarithmic modification. The parameter $w$ is tightly constrained to
$w=-0.7401^{+0.0367}_{-0.0385}$, indicating a well-localized posterior for the effective Herglotz-sector closure. The Herglotz field amplitude $\Phi_0$ is pushed close to its prior boundary, peaking near $\Phi_0 = 0.9861^{+0.0103}_{-0.0189}$, which may signal either a genuine preference for large $\Phi_0$ or a prior-driven feature that would benefit from an extended prior range in future analyses. The sound
horizon is constrained to $r_d = 148.1230^{+2.5803}_{-2.4834}~\mathrm{Mpc}$  \cite{Planck:2018vyg}, in good
agreement with both early-Universe (CMB-calibrated) and BAO-only determinations.\\

\begin{figure}
    \centering
    \includegraphics[width=1.0\linewidth]{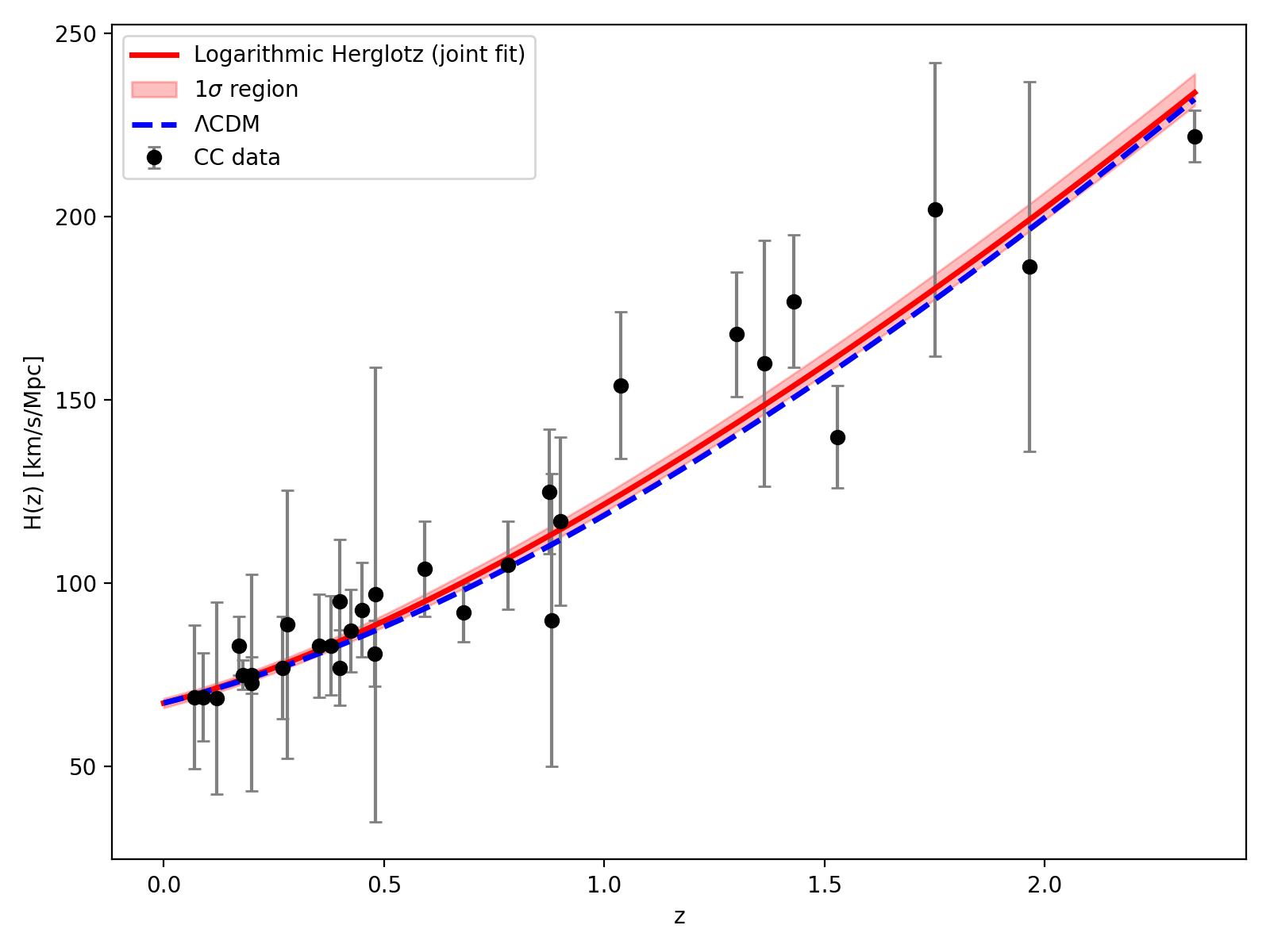}
    \caption{Hubble expansion history from the joint CC, Union3, and DESI DR2 BAO analysis. The solid curve shows the best-fitting logarithmic Herglotz model, with the shaded region denoting the $1\sigma$ posterior interval. The CC measurements are shown for comparison, together with the flat $\Lambda$CDM reference model for $\Omega_{m0}=0.30$.}
    \label{fig:combined_Hz}
\end{figure}

Figure~\ref{fig:combined_Hz} shows the best-fit $H(z)$ curve (with $1\sigma$ posterior-predictive band) from the joint fit against the cosmic
chronometer data points, together with a reference flat $\Lambda$CDM curve
at $\Omega_{m,0}=0.30$. The joint fit reproduces the CC measurements well
across the full redshift range while remaining simultaneously consistent
with the Union3 and DESI DR2 BAO data.

\begin{table*}[t]
\centering
\caption{Marginalized posterior constraints for the logarithmic Herglotz
$f(R,T)$ gravity model. For the BAO analysis, $r_d$ is included as a
nuisance parameter associated with the sound horizon at the drag epoch.}
\label{tab:parameter_constraints}
\renewcommand{\arraystretch}{1.4}
\begin{tabular}{lccccc}
\hline\hline
Dataset
& $H_0$
& $B$
& $w$
& $\Phi_0$
& $r_d$ \\
\hline
CC
& $65.3817^{+3.3622}_{-3.2932}$
& $-0.1099^{+0.1169}_{-0.2014}$
& $-0.7690^{+0.1534}_{-0.1962}$
& $0.9489^{+0.0379}_{-0.0672}$
& -- \\

BAO
& $73.9432^{+13.3087}_{-10.2559}$
& $-0.0718^{+0.0618}_{-0.1075}$
& $-0.8053^{+0.0593}_{-0.0706}$
& $0.9808^{+0.0137}_{-0.0232}$
& $137.5144^{+22.1771}_{-21.5214}$ \\

Union3
& $72.3953^{+3.0495}_{-2.8837}$
& $-0.6633^{+0.4994}_{-0.6160}$
& $-0.7356^{+0.0650}_{-0.0976}$
& $0.8765^{+0.0866}_{-0.1208}$
& -- \\

CC + Union3 + BAO
& $67.3488^{+1.2937}_{-1.2868}$
& $-0.0796^{+0.0628}_{-0.0933}$
& $-0.7401^{+0.0367}_{-0.0385}$
& $0.9861^{+0.0103}_{-0.0189}$
& $148.1230^{+2.5803}_{-2.4834}$ \\
\hline\hline
\end{tabular}
\end{table*}

\section{Diagnostic analysis}
\label{sec:Diagnostics}
\subsection{Evolution of the Deceleration Parameter}
\label{subsec:deceleration}

A useful diagnostic for investigating the dynamical evolution of the
Universe is the deceleration parameter,
\begin{equation}
q(z)
=
-1-\frac{\dot{H}}{H^2}.
\label{eq:q_definition}
\end{equation}
Using
\begin{equation}
\frac{dz}{dt}=-(1+z)H,
\end{equation}
the deceleration parameter can equivalently be expressed in terms of
the redshift derivative of the Hubble parameter as
\begin{equation}
q(z)
=
(1+z)\frac{1}{H(z)}
\frac{dH(z)}{dz}-1.
\label{eq:q_redshift}
\end{equation}
Since $H(z)=H_0h(z)$, the normalization \(H_0\) cancels from this expression, yielding
\begin{equation}
q(z)
=
(1+z)\frac{h'(z)}{h(z)}-1.
\label{eq:q_h}
\end{equation}
For the logarithmic Herglotz model, $h(z)$ does not admit a simple
closed-form expression and is therefore obtained by numerically
integrating the background field equations. The derivative $h'(z)$
is evaluated directly from the same system of differential equations,
ensuring that the deceleration parameter is calculated consistently
with the background solution used in the observational analysis.\\

\begin{figure}
    \centering
    \includegraphics[width=1.0\linewidth]{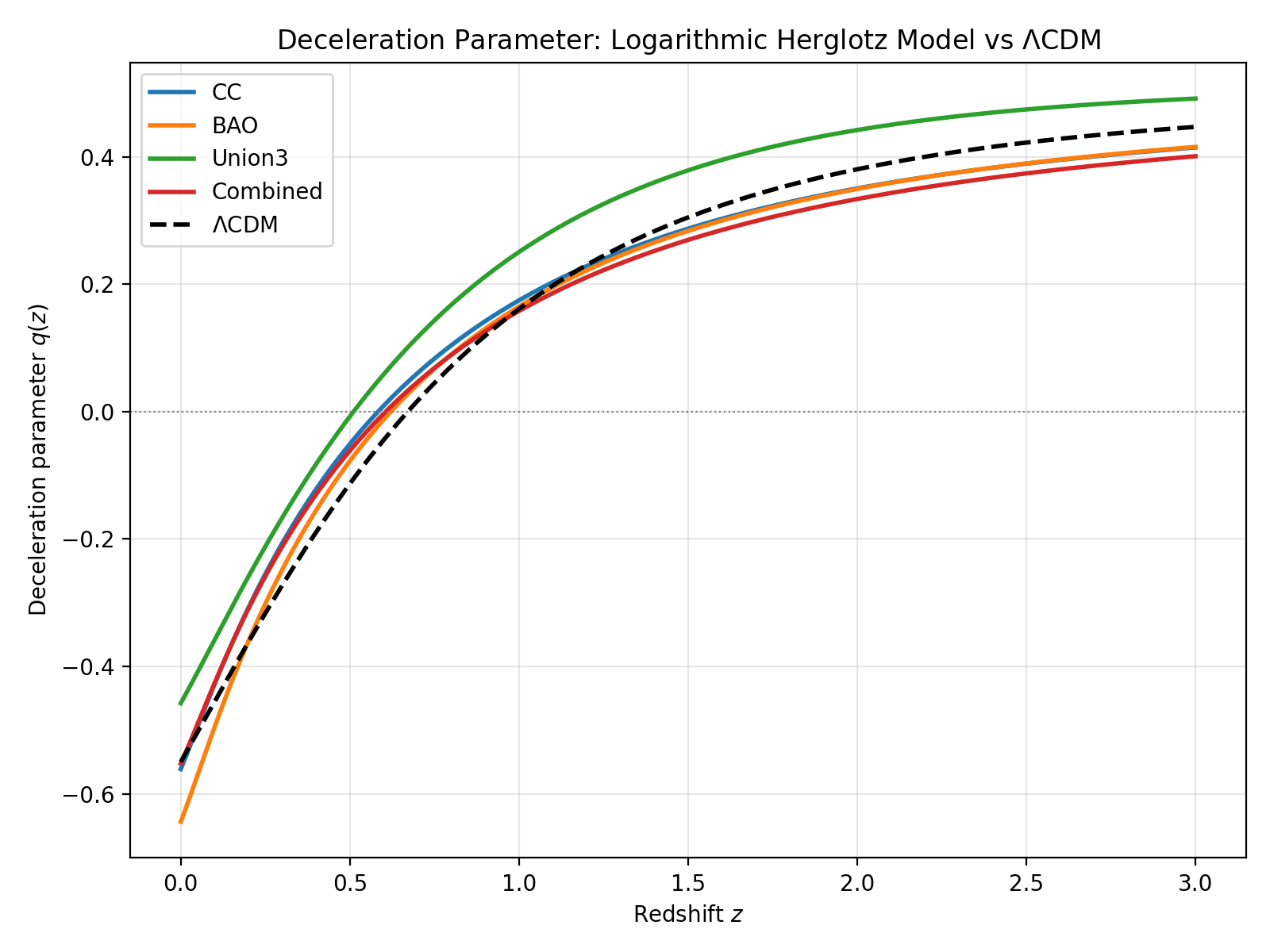}
    \caption{Evolution of the deceleration parameter $q(z)$ for the logarithmic Herglotz model using the posterior central values from the CC, BAO, Union3, and combined CC+Union3+DESI DR2 BAO analyses. The dashed curve shows the flat $\Lambda$CDM prediction for $\Omega_{m0}=0.30$, while the dotted line indicates $q(z)=0$.}
    \label{fig:qz_comparison}
\end{figure}

The evolution of $q(z)$ obtained using the marginalized posterior
central values of the model parameters from the CC, BAO, Union3, and
combined analyses is shown in Fig.~\ref{fig:qz_comparison}. For
comparison, the corresponding prediction of a spatially flat
$\Lambda$CDM model with $\Omega_{m0}=0.30$ \cite{Planck:2018vyg} is also displayed.

\subsection{$Om(z)$ Diagnostic}
\label{subsec:om_diagnostic}

The expansion history can be further characterized through the
$Om(z)$ diagnostic \cite{Sahni:2008xx}, which provides a useful kinematic comparison with the standard $\Lambda$CDM cosmology. It is
defined in terms of the normalized Hubble parameter
$E(z)=H(z)/H_0$ as
\begin{equation}
Om(z)
=
\frac{E^2(z)-1}{(1+z)^3-1},
\qquad
E(z)\equiv \frac{H(z)}{H_0}.
\label{eq:om_definition}
\end{equation}
For a spatially flat $\Lambda$CDM cosmology, $E^2(z)=\Omega_{m0}(1+z)^3+(1-\Omega_{m0})$,
and Eq.~\eqref{eq:om_definition} reduces identically to
$Om(z)\equiv\Omega_{m0}$ for all $z$.\\

For the logarithmic Herglotz model, the normalized expansion function
$E(z)=h(z)$ is obtained directly from the numerical solution of the
background field equations. Since $h(0)=1$, the $Om(z)$ diagnostic can
be constructed without an explicit dependence on the normalization
$H_0$. This allows the evolution of $Om(z)$ to be examined directly
for the posterior central values obtained from the individual CC,
BAO, and Union3 analyses, as well as from the combined observational
constraint.\\

\begin{figure}
    \centering
    \includegraphics[width=1.0\linewidth]{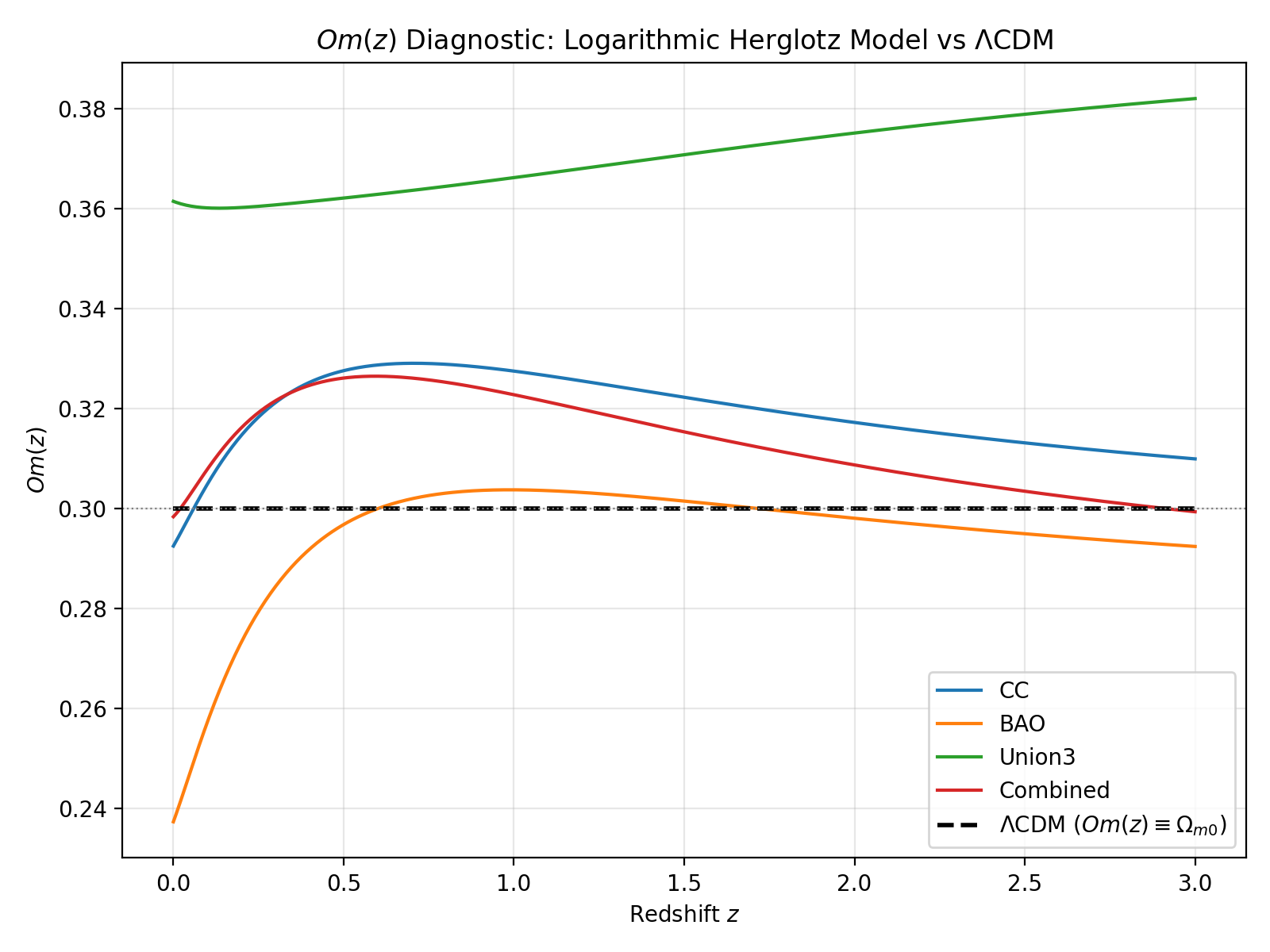}
    \caption{Evolution of the $Om(z)$ diagnostic for the logarithmic
Herglotz model using the posterior central values from the CC, BAO,
Union3, and combined CC+Union3+DESI DR2 BAO analyses. The horizontal
dotted line denotes the $\Lambda$CDM reference value
$\Omega_{m0}=0.30$, for which $Om(z)=\Omega_{m0}$ is constant.}
    \label{fig:Omz_comparison}
\end{figure}

Figure~\ref{fig:Omz_comparison} shows the resulting evolution of
$Om(z)$ over the redshift interval $0\leq z\leq3$. The horizontal
dotted line corresponds to $\Omega_{m0}=0.30$, while the dashed curve
shows the flat $\Lambda$CDM prediction. In contrast to $\Lambda$CDM,
for which $Om(z)$ remains exactly constant, all the logarithmic
Herglotz solutions exhibit a non-trivial redshift dependence. The
magnitude and direction of this evolution depend on the observational
dataset used to constrain the model, reflecting the different
parameter combinations preferred by the individual probes.

\subsection{Effective Equation of State}
\label{subsec:weff}

A useful kinematic characterization of the background expansion is
provided by the effective equation-of-state parameter. In the present
framework, the parameter $w$ appearing in the logarithmic Herglotz
field equations should not be identified with the equation of state
of the physical matter component, which is assumed to be pressureless,
$p_m=0$. Instead, $w$ characterizes the effective Herglotz sector
through the closure relation introduced in the background dynamics.
To characterize the overall expansion history, we define
\begin{equation}
w_{\rm eff}(z)
=
\frac{2q(z)-1}{3},
\label{eq:weff_definition}
\end{equation}
where $q(z)$ denotes the deceleration parameter. This quantity
provides an effective-fluid representation of the total background
expansion and does not correspond to an independently conserved dark
energy component.\\

For the logarithmic Herglotz model, the deceleration parameter is
obtained directly from the normalized Hubble function,
\begin{equation}
q(z)
=
(1+z)\frac{1}{h(z)}
\frac{dh(z)}{dz}-1,
\qquad
h(z)=\frac{H(z)}{H_0},
\label{eq:q_h_weff}
\end{equation}
where $h(z)$ is determined by numerically integrating the coupled
background equations. Substitution of the resulting $q(z)$ into
Eq.~\eqref{eq:weff_definition} therefore yields the effective
equation of state directly from the reconstructed expansion history.
No additional parametrization of $w_{\rm eff}(z)$ is introduced.\\

\begin{figure}
    \centering
    \includegraphics[width=1.0\linewidth]{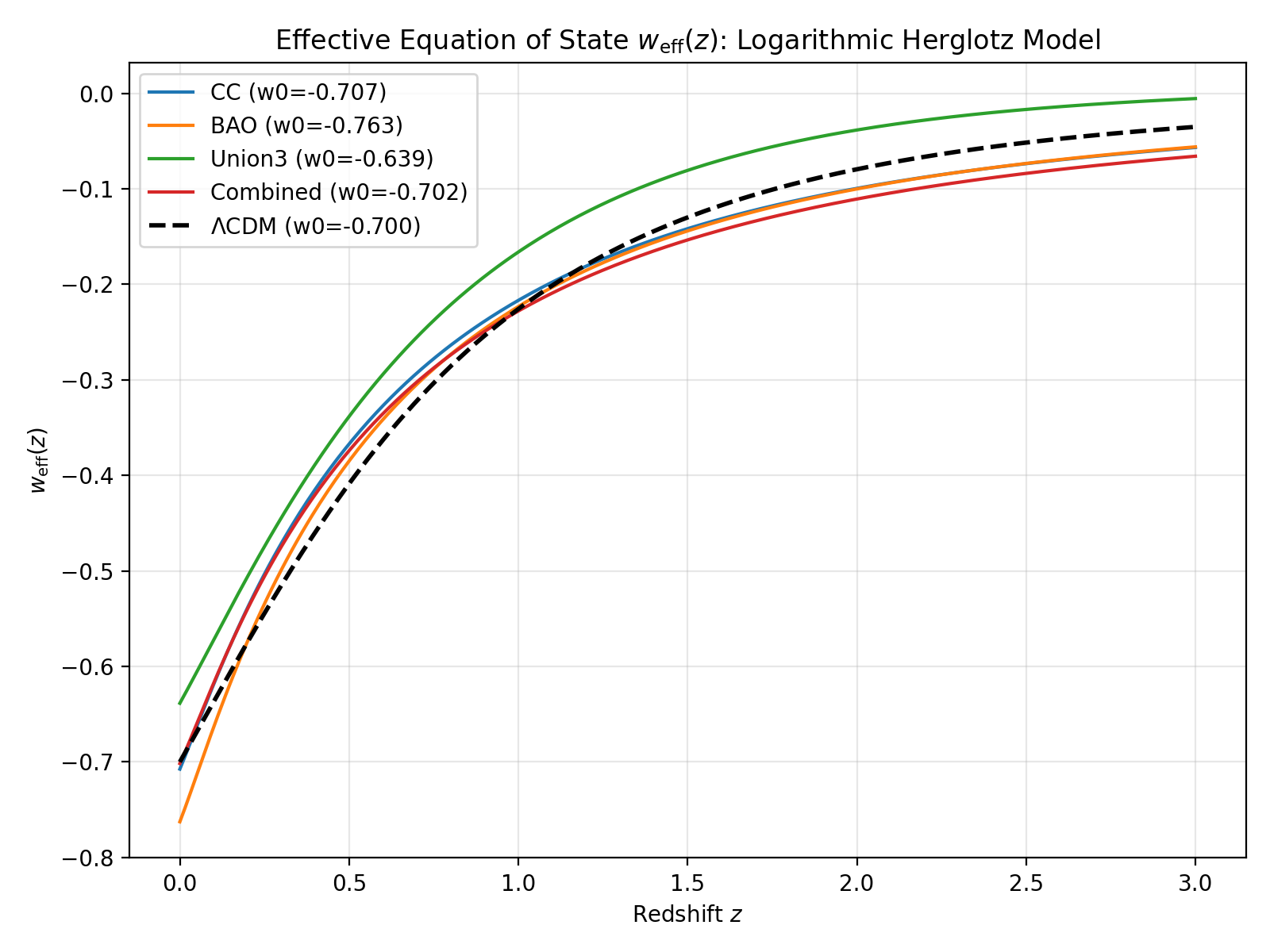}
    \caption{Redshift evolution of the effective equation-of-state parameter
$w_{\rm eff}(z)$ for the logarithmic Herglotz model using the posterior
central values from the CC, BAO, Union3, and combined analyses. The dashed
curve shows the flat $\Lambda$CDM prediction for $\Omega_{m0}=0.30$.
The present-day values are indicated in the legend.}
    \label{fig:weff_comparison}
\end{figure}

Figure~\ref{fig:weff_comparison} shows the evolution of
$w_{\rm eff}(z)$ obtained using the posterior central values from the
CC, BAO, Union3, and combined analyses. The corresponding
present-day values are
\begin{equation}
w_{\rm eff,0}\simeq
-0.707,\quad
-0.763,\quad
-0.639,\quad
-0.702,
\end{equation}
for the CC, BAO, Union3, and combined datasets, respectively.
The differences arise from the distinct background solutions
associated with the parameter constraints obtained from the individual
observational combinations.\\

For comparison, in a spatially flat $\Lambda$CDM cosmology containing
pressureless matter and a cosmological constant, the corresponding
effective equation of state is
\begin{equation}
w_{\rm eff}^{\Lambda{\rm CDM}}(z)
=
-\frac{1-\Omega_{m0}}
{\Omega_{m0}(1+z)^3+1-\Omega_{m0}}.
\label{eq:weff_lcdm}
\end{equation}
For the reference value $\Omega_{m0}=0.30$, this gives
\begin{equation}
w_{\rm eff}^{\Lambda{\rm CDM}}(0)=-0.70,
\end{equation}
and approaches zero at high redshift as the matter contribution
becomes increasingly dominant. The corresponding $\Lambda$CDM
prediction is shown by the dashed curve in
Fig.~\ref{fig:weff_comparison}.\\

The logarithmic Herglotz solutions exhibit the expected transition
from a negative effective equation of state at low redshift toward
values closer to zero at higher redshift. The magnitude of this
evolution depends on the observational dataset used to constrain the
model. In particular, the Union3-constrained solution remains
comparatively less negative over a substantial part of the redshift
interval, whereas the combined solution remains closer to the CC and
BAO reconstructions. These differences reflect the sensitivity of the
reconstructed expansion history to the observational combination used
in the parameter inference.

It is important to emphasize that the quantity $w_{\rm eff}(z)$ in
Eq.~\eqref{eq:weff_definition} is distinct from the constant
parameter $w$ entering the logarithmic Herglotz background equations.
The latter is a parameter associated with the effective Herglotz
sector and is introduced through its closure relation, whereas
$w_{\rm eff}(z)$ is a kinematic quantity reconstructed from the total
background expansion through $q(z)$. Consequently, the evolution of
$w_{\rm eff}(z)$ should be interpreted as a diagnostic of the
cosmological expansion history rather than as a direct determination
of the microscopic equation of state of the physical matter
component or of an independently conserved dark-energy fluid.

\subsection{Statefinder Diagnostics}
\label{subsec:statefinder}

To further investigate the dynamical behaviour of the logarithmic
Herglotz model, we employ the statefinder diagnostic introduced in
the $\{r,s\}$ parameter space. These geometrical diagnostics are
constructed directly from the deceleration parameter and provide
additional information about the higher-order evolution of the
cosmic expansion.\\

The statefinder parameters are defined as
\begin{equation}
r(z)
=
q(z)+2q^2(z)+(1+z)\frac{dq(z)}{dz},
\label{eq:statefinder_r}
\end{equation}
and
\begin{equation}
s(z)
=
\frac{r(z)-1}
{3\left[q(z)-\frac{1}{2}\right]}.
\label{eq:statefinder_s}
\end{equation}
The deceleration parameter entering these expressions is obtained
from the numerical solution of the background equations of the
logarithmic Herglotz model,
\begin{equation}
q(z)
=
(1+z)\frac{1}{h(z)}
\frac{dh(z)}{dz}-1,
\qquad
h(z)=\frac{H(z)}{H_0}.
\label{eq:q_statefinder}
\end{equation}
Since the logarithmic Herglotz model does not provide a closed-form
expression for $H(z)$, the quantities $q(z)$ and $dq/dz$ are evaluated
numerically from the same background solution employed in the
observational analysis.\\

\begin{figure}
    \centering
    \includegraphics[width=1.0\linewidth]{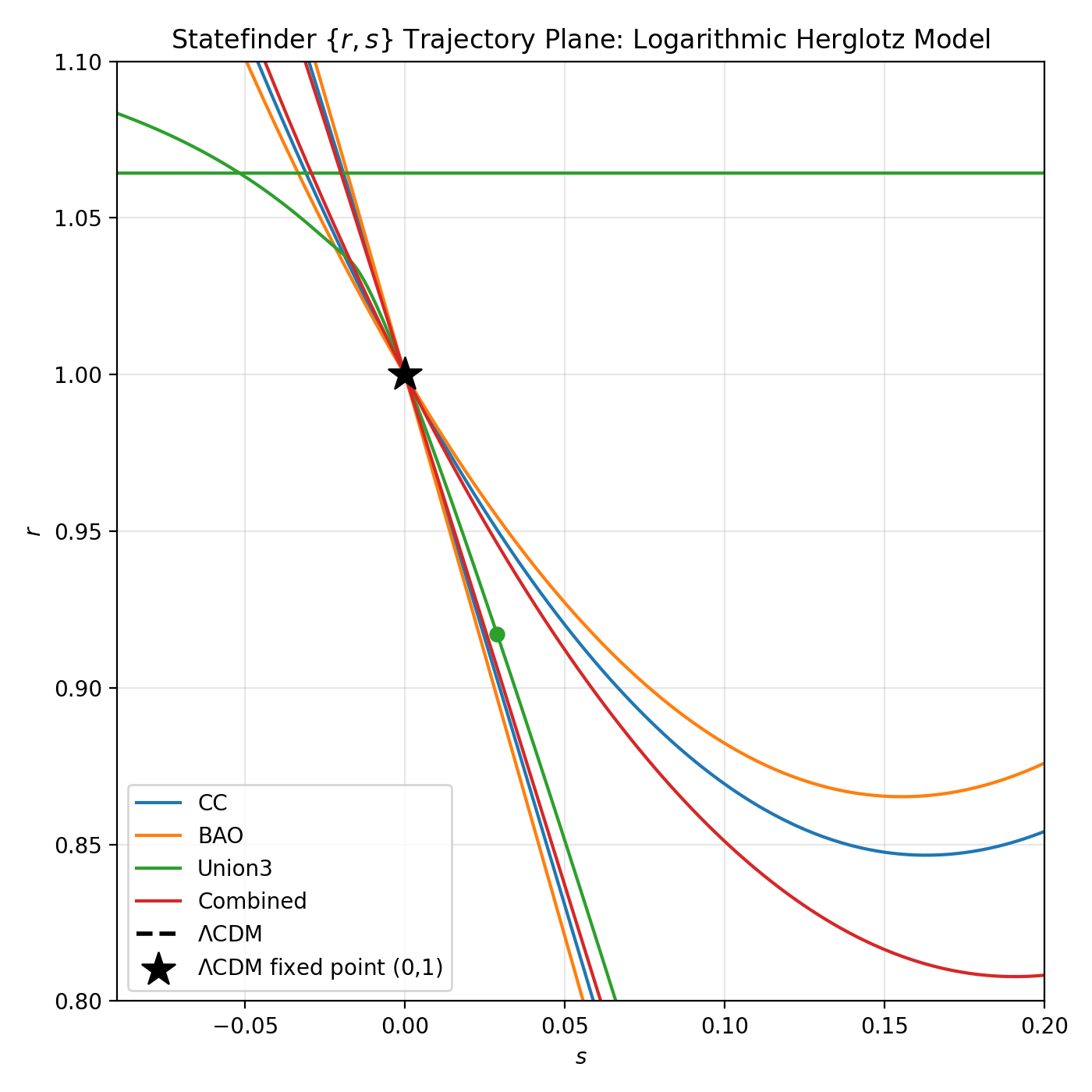}
    \caption{Statefinder $\{r,s\}$ trajectories of the logarithmic Herglotz model for the CC, BAO, Union3, and combined observational constraints. The black star marks the $\Lambda$CDM fixed point $(s,r)=(0,1)$. The trajectories deviate from this point as the redshift increases, with the evolution depending on the observational constraint.}
    \label{fig:statefinder_rs}
\end{figure}

Figure~\ref{fig:statefinder_rs} shows the trajectories of the
logarithmic Herglotz model in the $\{s,r\}$ plane obtained using the
posterior central values from the CC, BAO, Union3, and combined
datasets. The black star marks the
$\Lambda$CDM fixed point,
\begin{equation}
(s,r)_{\Lambda{\rm CDM}}=(0,1).
\label{eq:lcdm_fixedpoint}
\end{equation}
For the $\Lambda$CDM model, the statefinder parameters remain at this
fixed point throughout the cosmic evolution. Consequently, the
$\Lambda$CDM trajectory in the statefinder plane reduces to a single
point, represented by the black star in the figure.\\

The logarithmic Herglotz trajectories pass through the vicinity of
the $\Lambda$CDM fixed point at the present epoch, while subsequently
departing from it as the expansion history evolves. The different
observational datasets lead to distinct trajectories, reflecting the different parameter values preferred by the
corresponding observational constraints. In particular,
the BAO and CC solutions exhibit relatively similar trajectories,
whereas the Union3 and combined constraints produce visibly
different evolutionary paths in the statefinder plane.\\

The present-day locations of the trajectories provide a useful
geometrical measure of the deviation of the logarithmic Herglotz
model from the $\Lambda$CDM fixed point. A model that approaches
$(s,r)=(0,1)$ at the present epoch possesses an expansion history
that is correspondingly close to the $\Lambda$CDM prediction in this
kinematic diagnostic. The trajectories obtained here remain close to
the fixed point in the vicinity of the present epoch but evolve away
from it with increasing redshift, indicating that the logarithmic Herglotz model does not yield an
exactly $\Lambda$CDM-like trajectory in the statefinder plane.\\

The statefinder analysis is complementary to the $q(z)$,
$Om(z)$, and $w_{\rm eff}(z)$ diagnostics. While $q(z)$ characterizes
the acceleration or deceleration of the Universe, $Om(z)$ probes the
redshift dependence of the effective matter content, and
$w_{\rm eff}(z)$ provides an effective-fluid description of the
expansion, the $\{r,s\}$ plane incorporates higher-order information
through the derivative $dq/dz$. The combined use of these diagnostics therefore provides a
complementary characterization of the background dynamics of the
logarithmic Herglotz model.

\section{Conclusion}
\label{sec:conclusion}

In this work, we have investigated the cosmological dynamics and
observational constraints of a logarithmic Herglotz-type $f(R,T)$
gravity model. The Herglotz variational framework introduces an
additional one-form associated with non-conservative gravitational
dynamics, while the dependence on the energy--momentum trace $T$
provides a direct matter--geometry coupling. Starting from the
general Herglotz-type $f(R,T)$ field equations, we considered a
spatially flat FLRW background and adopted the logarithmic form
\begin{equation}
    f(R,T)
    =
    R+\beta\ln\left(-\frac{T}{T_0}\right).
\end{equation}
For pressureless matter, the resulting cosmological equations were
written as a closed system of first-order differential equations for
the dimensionless Hubble parameter, the Herglotz variable, and the
normalized matter density. The system was solved numerically to
reconstruct the background expansion history.\\

The model was constrained using three complementary late-time
observational probes: cosmic chronometer (CC) measurements, DESI
DR2 baryon acoustic oscillation (BAO) measurements, and the Union3
Type Ia supernova compilation. The model parameters were constrained
independently for each dataset using an affine-invariant Markov Chain
Monte Carlo analysis. For the BAO analysis, the sound horizon at the
drag epoch was treated as a free parameter and the full covariance
matrix was incorporated. A joint analysis of the CC, BAO, and Union3
datasets was then performed to obtain combined constraints on the
model parameters.\\

The observationally constrained solutions yield a transition from
decelerated expansion at higher redshift to accelerated expansion at
late times. The reconstructed expansion histories show dataset
dependent differences, reflecting the distinct constraints provided
by the individual observational probes. The combined analysis
provides a parameter realization constrained simultaneously by the
three datasets.\\

Several complementary diagnostics were employed to further examine
the reconstructed expansion history. The $Om(z)$ diagnostic exhibits
a non-trivial redshift dependence, in contrast to the constant value
obtained in flat $\Lambda$CDM. For the combined constraint, the
reconstructed evolution remains close to the $\Lambda$CDM reference
over the redshift range considered, while retaining a distinct
redshift dependence.\\

The effective equation-of-state parameter provides a complementary
kinematic description of the expansion. The present-day values
obtained from the CC, BAO, Union3, and combined constraints are
approximately $-0.707$, $-0.763$, $-0.639$, and $-0.702$,
respectively. The effective equation of state becomes less negative
with increasing redshift, consistent with the transition from
late-time acceleration toward an earlier matter-dominated regime.
Here, $w_{\rm eff}(z)$ is a kinematic quantity reconstructed from the
background expansion and is distinct from the constant parameter $w$
appearing in the underlying Herglotz cosmological equations.\\

The statefinder analysis further characterizes the higher-order
evolution of the expansion. The logarithmic Herglotz trajectories
remain close to the $\Lambda$CDM fixed point $(s,r)=(0,1)$ around the
present epoch and deviate from it as the redshift increases. The
differences among the trajectories obtained from the individual
datasets reflect the corresponding differences in the reconstructed
background solutions.\\

Overall, the logarithmic Herglotz $f(R,T)$ model provides a viable
description of the late-time background expansion within the
observational framework considered here. The combined CC, DESI DR2
BAO, and Union3 constraints, together with the $q(z)$, $Om(z)$,
$w_{\rm eff}(z)$ and statefinder analyses,
demonstrate that the model can reproduce the principal features of
the reconstructed late-time expansion while allowing departures from
the exact $\Lambda$CDM evolution. The present study is restricted to
the homogeneous and isotropic background level. Further investigation
of perturbation growth, structure formation, and additional
observational probes will therefore be important for testing the
model beyond the background cosmology considered here.

\bibliographystyle{unsrtnat}
\balance
\bibliography{reference}

\end{document}